\documentclass{sig-alternate-2013}

\usepackage[pass]{geometry}

\usepackage{endnotes}
\usepackage{times}
\usepackage[T1]{fontenc}
\usepackage{float}
\usepackage{graphicx}
\usepackage{array}
\usepackage{url}
\usepackage{comment}
\usepackage{framed}
\usepackage{multirow}
\usepackage{hyperref}
\usepackage{listings}

\usepackage{amsmath,amsthm,amssymb,latexsym, pifont}
\usepackage[usenames]{color}
\usepackage{ifsym}
\usepackage{wasysym}
\usepackage{booktabs}
\usepackage{mdwlist}
\usepackage{caption}
\usepackage{subcaption}
\usepackage{latexsym}
\usepackage{epstopdf}
\usepackage{color}
\usepackage{enumitem}
\usepackage{microtype}
\usepackage[ruled]{algorithm2e}
\usepackage{cite}
\usepackage{xspace}
\usepackage{xparse}
\usepackage{nccmath}

\newcommand{\xref}[1]{Section~\ref{#1}}
\newcommand{\cref}[1]{Chapter~\ref{#1}}

\newcommand{\fref}[1]{Fig.~\ref{#1}}

\newcommand{\first}{\emph{(i)}~}
\newcommand{\second}{\emph{(ii)}~}
\newcommand{\third}{\emph{(iii)}~}
\newcommand{\fourth}{\emph{(iv)}~}

\newcommand{\ie}{i.e., \@}
\newcommand{\eg}{e.g., \@}

\definecolor{darkgreen}{rgb}{0,0.5,0}
\definecolor{brown}{rgb}{0.7,0.3,0}
\definecolor{darkblue}{rgb}{0,0,0.5}

\newcounter{fn1}
\newcounter{fn2}
\newcounter{fn3}
\newcounter{fn4}
\newcounter{fn5}
\theoremstyle{definition}

\let\underscore\_
\newcommand{\myunderscore}{\renewcommand{\_}{\underscore\hspace{0pt}}}
\myunderscore

\newcommand{\hide}[1]{}

\definecolor{darkblue}{rgb}{0,0,0.5}
\hypersetup{pdfborder=0 0 0,
    colorlinks=true,
    citecolor=darkblue,
    linkcolor=darkblue,
    urlcolor=darkblue
  }

\def\algMCF/{\ensuremath{\texttt{MinCostFlow}}}
\def\algMCA/{\ensuremath{\texttt{MinCostAssignment}}}
\def\algSP/{\ensuremath{\texttt{ShortestPath}}}
\def\algMSP/{\ensuremath{\texttt{MinAllShortestPath}}}

\def\algLP/{\ensuremath{\texttt{solveLP}}}
\def\algSolveSepSolve/{\ensuremath{\texttt{solveSeparateSolve}}}
\def\algAddLocalConstraints/{\ensuremath{\texttt{addConstraintsLocally}}}

\def\infeasibleLP/{\ensuremath{\texttt{infeasibleLP}}}
\def\objectiveLimit/{\ensuremath{\texttt{objectiveLimit}}}
\def\disableGlobalCutoff/{\ensuremath{\texttt{disableGlobalPrimalBound}}}

\def\NULL/{\ensuremath{\texttt{null}}}

\newfont{\mycrnotice}{ptmr8t at 7pt}
\newfont{\myconfname}{ptmri8t at 7pt}
\let\crnotice\mycrnotice
\let\confname\myconfname

\permission{}
\conferenceinfo{}{} 
\copyrightetc{}
\crdata{}

\begin{document}

\title{Investigating the Potential of the Inter-IXP Multigraph for the Provisioning of Guaranteed End-to-End Services}

\numberofauthors{3}
\author{
\alignauthor
Vasileios Kotronis,\\ Rowan Kl\"oti,\\ \mbox{Panagiotis Georgopoulos},\\ Bernhard Ager\\
\affaddr{ETH Zurich, Switzerland}\\ 
\alignauthor{Matthias Rost,\\ Stefan Schmid}\\
\affaddr{TU Berlin, Germany}\\
\alignauthor{\mbox{Xenofontas Dimitropoulos}}\\
\affaddr{FORTH, Greece}\\
\affaddr{ETH Zurich, Switzerland}
}

\maketitle

\begin{abstract}

In this work, we propose utilizing the rich connectivity between IXPs and ISPs for inter-domain
path stitching, supervised by centralized QoS brokers. In this
context, we highlight a novel abstraction of the Internet topology, \ie
the inter-IXP multigraph composed of IXPs and paths crossing the domains
of their shared member ISPs. This can potentially serve as a dense Internet-wide substrate 
for provisioning guaranteed end-to-end (e2e) services with high
path diversity and global IPv4 address space reach. We thus map 
the IXP multigraph, evaluate its potential, and
introduce a rich algorithmic framework for path stitching on
such graph structures.

\end{abstract}

\category{C.2.2}{Network Protocols}{Routing Protocols}
\keywords{Internet Exchange Point;~QoS;~Embedding;~EuroIX}

\section{Introduction}

Modern Internet applications, from HD video-conferencing to 
telesurgery and remote control of power-plants, 
pose increasing demands on network latency, bandwidth and availability. 
Presently, ISPs are able to provide certain QoS guarantees only in
intra-domain settings based on technologies such as leased circuits and VPN
tunnels, \eg over MPLS. 
Such services provide an important revenue
stream for ISPs. However, despite several research and standardization 
efforts, providing QoS guarantees at the inter-domain
level has seen very limited success so far.

Centralized inter-domain path brokers have been explored 
in the literature~\cite{RouteSource,MINT,QoSIPMPLS} 
and the industry~\cite{GEANT}, as an approach 
to support the requirements of demanding applications across domains.
In these setups, ISPs provide QoS-enabled pathlets~\cite{Pathlet}, which are stitched together 
by an inter-domain routing mediator, \eg a bandwidth broker mediating the exchange 
of transit bandwidth between ISPs~\cite{MINT}. 
This work explores logically centralized inter-domain QoS routing
mediators for the provisioning of inter-domain path guarantees
in light of ongoing changes in the Internet ecosystem. In particular,
\first the Internet is becoming denser and more flat~\cite{FlatInternet}
due to richer interconnectivity at Internet eXchange Points (IXPs)~\cite{EuroIX}, and
\second there is a gradual paradigm shift towards network virtualization and 
Software Defined Networking (SDN), also in the context of IXPs~\cite{SDX-SIGCOMM}.
Our main proposal is that a rich inter-IXP overlay fabric opens interesting 
traffic engineering flexibilities which can be exploited, \eg for inter-domain QoS. 

\section{CXPs: Path Brokers over IXPs}
\label{sec:cxps}

\begin{figure}[t]
\begin{center}
  \includegraphics[width=1.0\columnwidth]{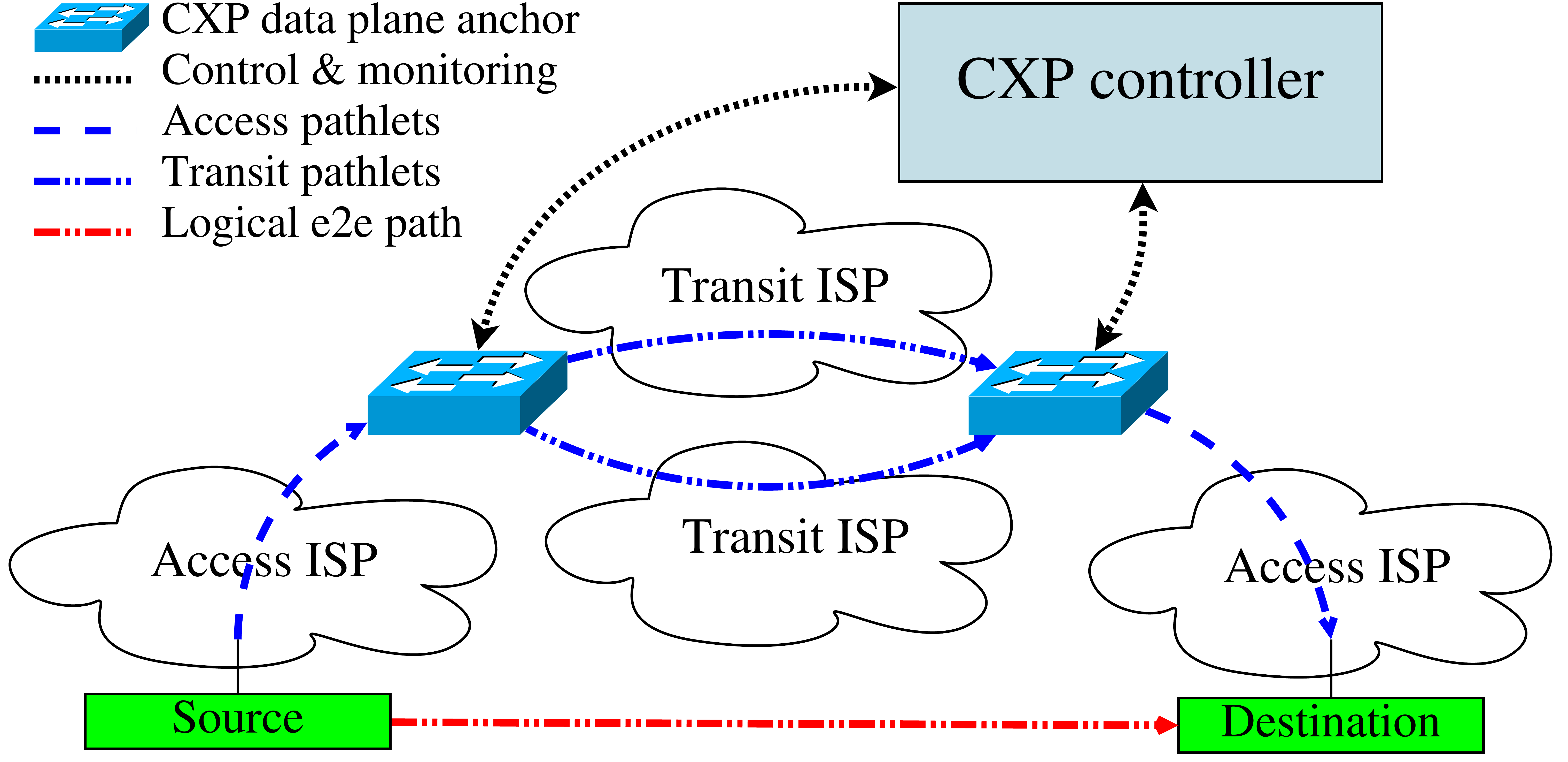}
	  \caption{The CXP stitches QoS-enabled e2e paths.}
	\label{fig:cxp-concept}
\vspace{-25pt}
\end{center}
\end{figure}

We propose performing inter-domain QoS routing over a novel abstraction of the
Internet topology, in which vertices are IXPs and edges are virtual
links connecting two IXPs over an ISP. We call this abstraction the
\emph{IXP multigraph} because two IXPs can be connected with multiple
edges over different ISPs. The choice of IXPs as  switching points 
exploits their rich connectivity to geo-diverse ISPs, enabling high path diversity and global client
reach with a deployment of only of a few well-placed switching points as
we show in \xref{sec:multigraph}. We
call the routing brokers and controllers that are deployed over IXP multigraphs \emph{Control eXchange Points (CXPs)}.
CXPs stitch pathlets~\cite{Pathlet} across multiple
administrative domains to construct global paths. The pathlets are provided by ISPs and are annotated
with specific properties, such as bandwidth and latency guarantees.
The incentive for ISPs to provide pathlets
is the revenue generated by their use for e2e services. 
As shown in \fref{fig:cxp-concept}, the ISPs of the source and
the destination offer access pathlets to connect to ISP-adjacent data
plane anchors, \ie programmable CXP switching points, while the intermediary ISPs offer transit pathlets between those points.

A CXP performs the following tasks: it \first handles new requests
for QoS-enabled paths (admission control), \second computes and sets up suitable paths
(embeddings), \third monitors pathlet availability and compliance
with QoS guarantees, and \fourth performs re-embedding, if required.
A client negotiates her request directly with her access ISP, which
selects a suitable CXP for establishing the inter-domain route out
of a set of available CXPs. The ISP forwards the client's request to
the chosen CXP, which in turn computes a suitable e2e path. The
CXP reserves capacity on the selected pathlets and then configures
the respective data plane anchors. Accordingly, the client's ISP has
to configure its network such that the quality sensitive traffic is sent
via a pathlet to the correct data plane anchor. A CXP monitors the
bandwidth, latency and availability of a path for the duration of
the client's reservation. If the client's requirements are
violated or a pathlet becomes unavailable, the CXP chooses and
configures an alternate path for the affected part(s) of the traffic;
this can even be a ``hot-standby'' backup path carrying traffic duplicates.
Besides, the CXP may choose to better utilize the available pathlets
by re-embedding paths and defragmenting the substrate resources.
CXPs and ISPs may use MPLS, SDN or PCE mechanisms for the
execution of their associated tasks, such as Traffic Engineering (TE).

\section{Potential of IXP Multigraphs}
\label{sec:multigraph}

\begin{figure}[t]
\centering
  \begin{subfigure}[b]{0.49\columnwidth}
  \includegraphics[trim= 10mm 10mm 0mm 12mm,clip=true,width=1.0\columnwidth]{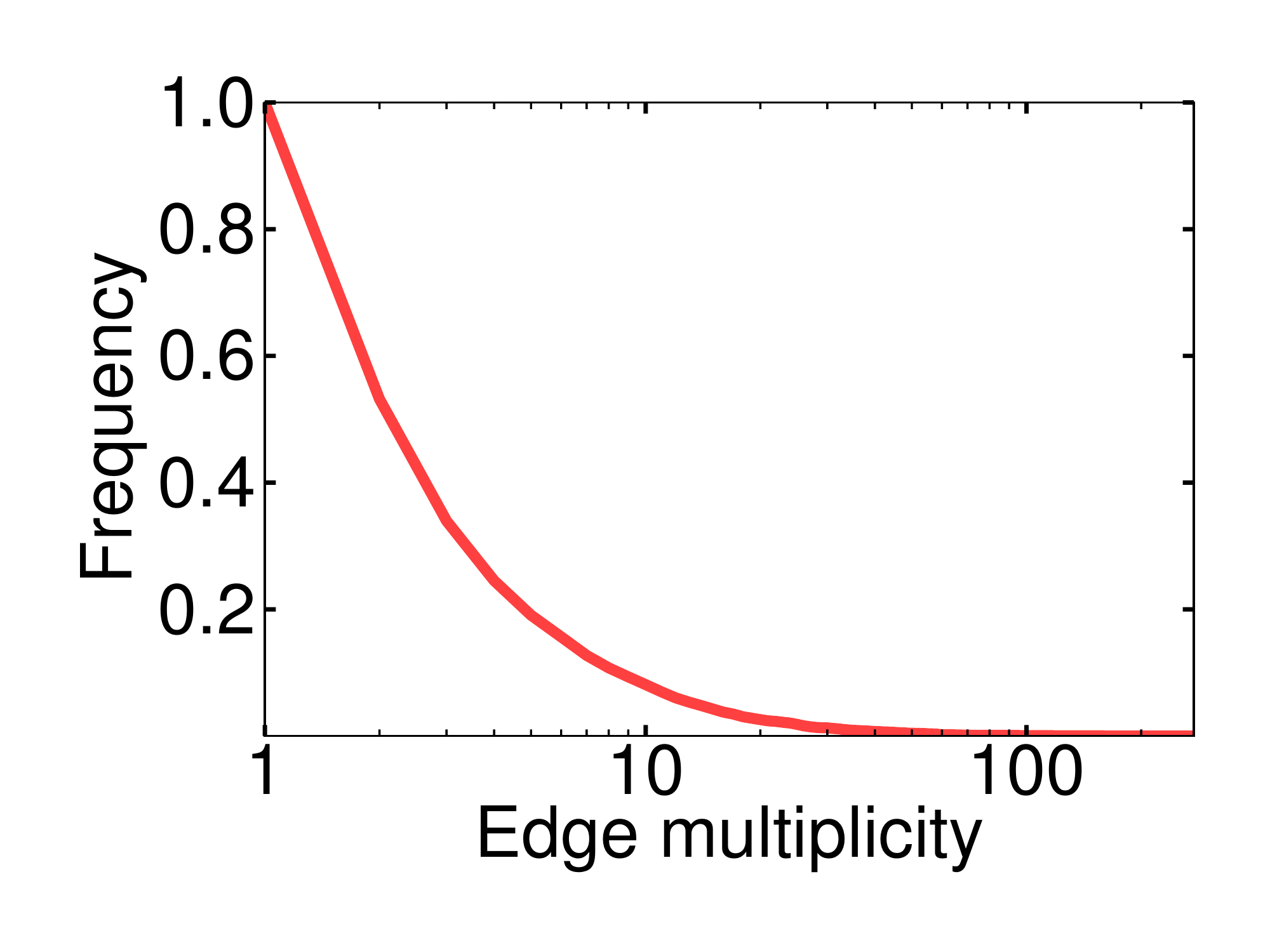}
\label{fig:common-as}
  \end{subfigure}
  \begin{subfigure}[b]{0.49\columnwidth}
  \includegraphics[trim= 10mm 10mm 0mm 12mm,clip=true,width=1.0\columnwidth]{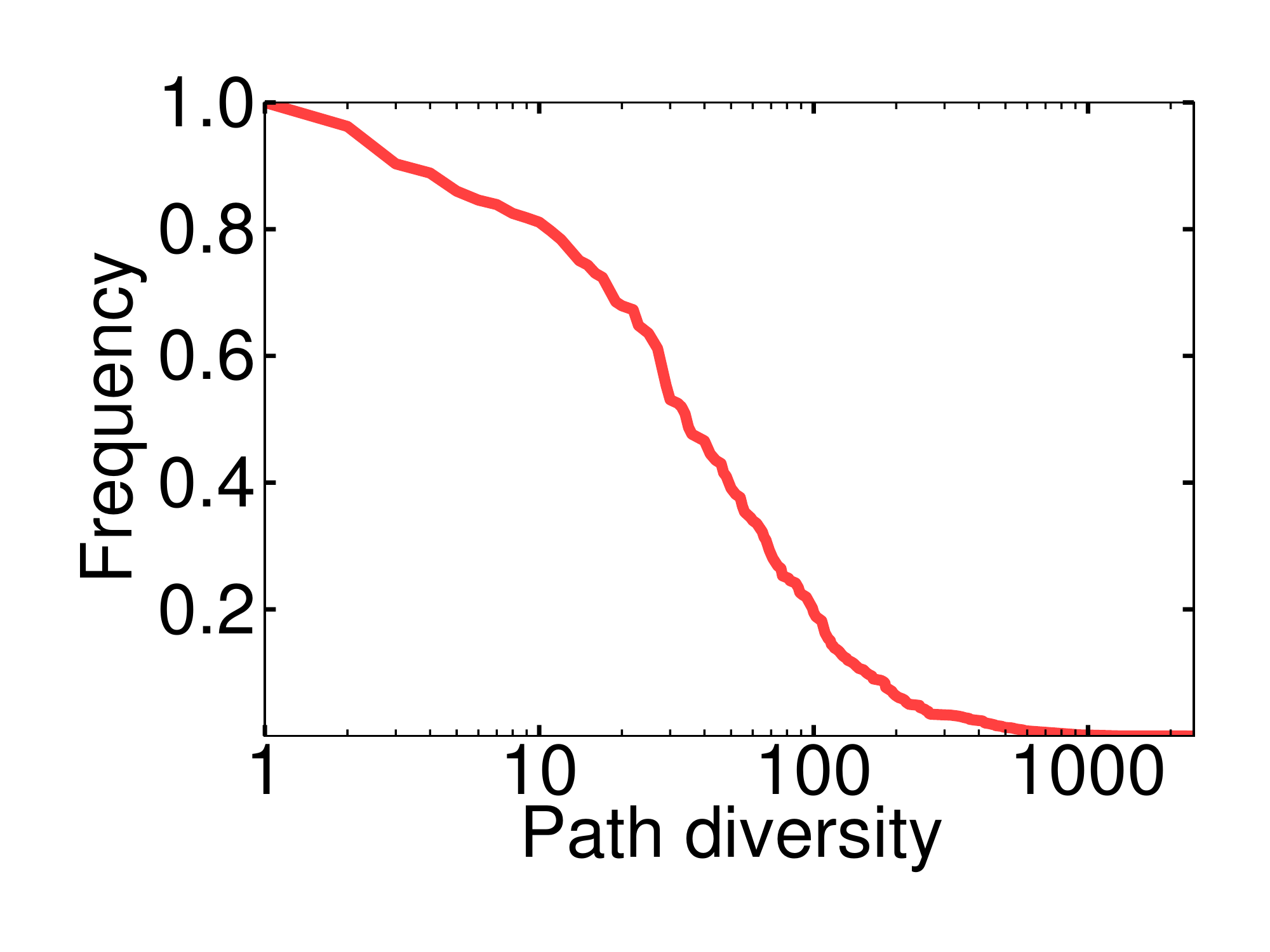}
\label{fig:path-div}
  \end{subfigure}
\vspace{-10pt}
\caption{CCDFs of key properties of the IXP multigraph}
\label{fig:empd-fig}
\vspace{-15pt}
\end{figure}

Motivated by the CXP concept, we analyze the properties of the resulting IXP multigraph
by extracting ISP membership data for 229 IXPs from the EuroIX dataset~\cite{EuroIX}.
We find \emph{49k}
inter-connections of type: $IXP_A, ISP_X, IXP_B$. These form a dense multigraph,
where a few pairs of IXPs are interconnected by hundreds of distinct ASes, each
of which is in a position to offer one or multiple pathlets between each pair. 
In \fref{fig:empd-fig}, we observe that the resulting pathlet-wise path diversity 
is one order of magnitude larger than the direct IXP 
connectivity, assuming that all the member ASes provide one
pathlet each per IXP pair.
In terms of AS-level path diversity, in our accompanying technical report~\cite{kotronis2015invtech}
we show that the application of generalized inter-domain routing policies can lead to an 
increase of up to \emph{29} times, as compared to classic valley-free practices~\cite{NoGlobalCoor}. 
More disjoint paths indicate potentially higher availability
for demanding applications. In contrast, classic BGP-based
routing is known to use a limited number of suboptimal inter-domain
paths~\cite{lumezanu2009triangle}, as they normally cross up to one peering link 
in an IXP.
Furthermore, in terms of IP address coverage, we discovered that even a small deployment
($\sim$5 IXP anchors) could directly cover a high fraction ($\sim$40\%)
of the Internet IPv4 address space. This increases
to $\sim$91\% of announced addresses if we
also consider the 1-hop customer cone of the IXP members, assuming access pathlets
over the member ISPs. This allows an initial deployment
of just a few IXPs to serve large sets of IPv4 addresses and
enables incremental adoption. Further use of private peering points
might selectively extend the graph and augment coverage, where required.

\section{Path Stitching Algorithms}
\label{sec:algos}

Given the interesting properties of the IXP-based multigraph, 
we devised algorithms which leverage 
its flexibility in the context of CXP-like QoS mediators. 
The problem is essentially a path embedding problem:
CXPs embed e2e paths on dense inter-IXP multigraphs, subject
to bandwidth and latency constraints. The objective is to exploit the rich path diversity in order 
to maximize the number of concurrently embeddable routes. These routes are requested
by the client IPv4 endpoints that connect to the multigraph via their access ISPs.
In our accompanying technical report~\cite{kotronis2015invtech}, we formally introduce 
the e2e routing problem considered in this work as 
the QoS Multigraph Routing Problem (QMRP), together with an optimal offline
formulation. This problem is complex
for several reasons. \first Requests from the large client base dynamically arrive 
over time in a non-predictable manner,
necessitating the use of online algorithms. \second While a
single suitable e2e path can be found in polynomial time, the IXP-based graph offers
many choices and requires to carefully
select which of the edges between two IXPs is used. In fact, in such multigraphs
the number of available edges can be 3 orders of magnitude larger than the
number of nodes (IXPs)~\cite{kotronis2015invtech}. \third The online
selection of e2e paths should reflect multiple conflicting high-level
objectives, namely accepting as many requests as possible, avoiding
the usage of scarce low-latency, high-bandwidth links, and preventing
resource fragmentation.

We thus propose a \emph{sample-select}
approach to heuristically tackle the QMRP in an online manner. In particular, given
the NP-hardness of the \emph{optimal path} calculations, we employ a sample-select
process, where in the first stage, a set of \emph{feasible} paths is
sampled (\ie generated) in polynomial time, and subsequently one of them is selected
for the actual embedding. The techniques for the sampling process may
range from modified Dijkstra-based approaches to random walks. 
Moreover, we propose a hybrid online-offline algorithm to support 
reconfigurations of pre-generated embeddings
in order to accommodate further online requests.
Using simulations based on the mapped
IXP multigraph, we show that our algorithms scale to the sizes of
the measured graphs and can serve diverse path request mixes.
The full set of online, offline and hybrid algorithmic variants,
together with the results and insights stemming from their application
on the EuroIX-based multigraph, can be found in our technical report~\cite{kotronis2015invtech}.
The software that we used is publicly available~\cite{cxp2015coderepo}.

\textbf{Acknowledgments.}
This work was partly funded by European Research Council Grant Agreement no. 338402.

\bibliographystyle{acm}
{\scriptsize
\bibliography{smetric061-kotronis}
}

\end{document}